\pdfoutput=1

\documentclass[11pt,twoside,a4paper,cmspaper,final,collab]{cms-tdr}
\def\svnVersion{418301P}\def\cmsCernNoTag{CERN-EP/2016-319}\def\cmsCernDate{\today}\def\cmsMessage{Published in Physics Letters B as \href{http://dx.doi.org/10.1016/j.physletb.2017.07.022}{\doi{10.1016/j.physletb.2017.07.022.}}}

\begin{document}\cmsNoteHeader{B2G-16-006}

\hyphenation{had-ron-i-za-tion}
\hyphenation{cal-or-i-me-ter}
\hyphenation{de-vices}
\RCS$Revision: 412076 $
\RCS$HeadURL: svn+ssh://alverson@svn.cern.ch/reps/tdr2/papers/B2G-16-006/trunk/B2G-16-006.tex $
\RCS$Id: B2G-16-006.tex 412076 2017-06-21 15:40:36Z alverson $
\newlength\cmsFigWidth
\ifthenelse{\boolean{cms@external}}{\setlength\cmsFigWidth{0.49\textwidth}}{\setlength\cmsFigWidth{0.6\textwidth}}
\ifthenelse{\boolean{cms@external}}{\providecommand{\cmsLeft}{top\xspace}}{\providecommand{\cmsLeft}{left\xspace}}
\ifthenelse{\boolean{cms@external}}{\providecommand{\cmsRight}{bottom\xspace}}{\providecommand{\cmsRight}{right\xspace}}

\providecommand{\MCatNLO}{{\MADGRAPH{}5\_a\MCATNLO}\xspace}
\newcommand{\Y} {\ensuremath{\mathrm{Y}}\xspace}
\newcommand{\T} {\ensuremath{\mathrm{T}}\xspace}

\cmsNoteHeader{B2G-16-006}
\title{Search for single production of vector-like quarks decaying into a \PQb quark
and a \PW~boson in proton-proton collisions at $\sqrt{s} = 13\TeV$}

\date{\today}
\abstract{
A search is presented for a heavy vector-like quark, decaying into a b quark and a W boson, which is produced singly in association with a light flavor quark and a b quark. The analysis is performed using a data sample of proton-proton collisions at a center-of-mass energy of $\sqrt{s} = 13\TeV$ collected at the LHC in 2015. The data set used in the analysis corresponds to an integrated luminosity of 2.3\fbinv. The search is carried out using events containing one electron or muon, at least one b-tagged jet with large transverse momentum, at least one jet in the forward region of the detector, and missing transverse momentum. No excess over the standard model prediction is observed. Upper limits are placed on the production cross section of heavy exotic quarks: a T quark with a charge of $2/3$, and a Y quark with a charge of $-4/3$. For Y quarks with coupling of 0.5 and ${\mathcal{B}}(\Y \to \PQb \PW) = 100\%$, the observed (expected) lower mass limits are 1.40 (1.0)\TeV. This is the most stringent limit to date on the single production of the Y vector-like quark.
}
\hypersetup{%
pdfauthor={CMS Collaboration},%
pdftitle={Search for the singly-produced vector-like partner of the top quark decaying into a W boson and a b quark},%
pdfsubject={CMS},%
pdfkeywords={CMS, physics}}
\maketitle

\section{Introduction}
\label{sec:intro}
The standard model (SM) of particle physics
has been exceptionally successful
in describing phenomena at the subatomic scale.
The observation of a Higgs boson with a mass of 125\GeV and with properties consistent with the SM
expectations~\cite{Atlas_Higgs, CMS_Higgs, Higgs_long} completed the SM.
However, in the absence of enormous order-dependent cancellations, also known as fine-tuning,
large SM quantum corrections would shift the bare Higgs boson mass to values far beyond the electroweak scale.
New physics is required to stabilize the Higgs boson mass naturally at the electroweak scale, i.e. without invoking
fine-tuning.

Many natural extensions of the SM have been proposed in recent decades.
Some of these models postulate the existence of vector-like quarks
(VLQs)~\cite{Lodone:2008yy,Perelstein:2003wd,Contino:2006nn}, which are colored fermions
with left- and right-handed chiral states both transforming in the same way under the gauge group
$\mathrm{SU(3)_{C}}\times \mathrm{SU(2)_{L}}\times \mathrm{U(1)_Y}$. The VLQs do not acquire masses through the Yukawa coupling to
the Higgs field,
and could cancel loop corrections from the SM top quark to the Higgs boson mass.

Searches for VLQs have already been performed
in various decay modes using proton-proton collisions at $\sqrt{s} = 8\TeV$. These searches were primarily focused on the pair production mechanism and they ruled out
VLQs with masses up to approximately 0.90\TeV~\cite{TPrime, BPrime, ATLAS-TPrime, ATLAS-BPrime}.
The VLQ  single production mechanism is coupling-dependent, and it could become the dominant contribution to the cross section at high VLQ masses.
The strength of the VLQ-b-W coupling can be approximately characterized by a single dimensionless parameter that varies from 0 to $\sqrt{2}$~\cite{DeSimone2013}, where the latter would correspond to a coupling of full electroweak strength.

\begin{figure}[htp]
\centering
\includegraphics[width=0.49\textwidth]{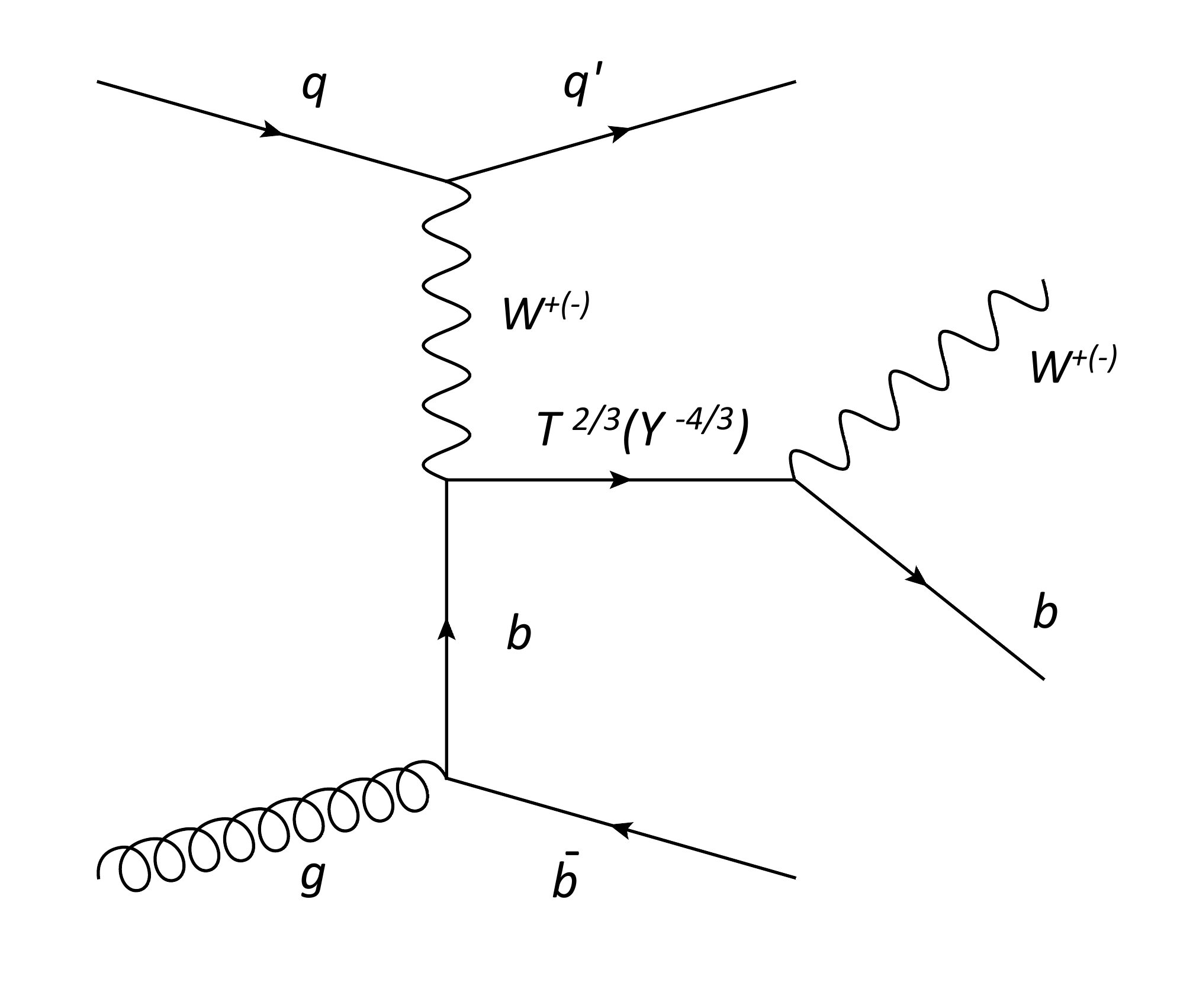}
\caption{Leading order Feynman diagram for singly produced Y or T quarks.
}
\label{fig:feydiagram}
\end{figure}

In this paper, we present a search for the single production of a heavy vector-like quark
that decays into a b quark and a W boson using the 2015 LHC data set. This
signature can arise from either a Y or a T quark with a charge of $-4/3$
or $2/3$, respectively, produced in association with a light flavor quark and a b quark.
The leading order Feynman diagram for Y and T quark production is shown in Fig.~\ref{fig:feydiagram}.
The outgoing light flavor quark $\PQq^\prime$ in the upper
part of the diagram produces a jet in the forward region of the detector,
which is a distinct signature of single production.

The Y quark is expected to decay with a branching fraction (${\mathcal{B}}$) of 100\% into
a b quark and a W boson~\cite{handbook}, while the T quark can also decay into tH and tZ via a flavor changing neutral current.
Searches with the 2015 LHC data set for single production of a vector-like T quark decaying to tH and tZ have
been performed by the CMS Collaboration~\cite{CMS_single_Ht_lep, CMS_single_Ht_had, CMS_single_Hz}.
If the T quark is a singlet, then it is expected to decay into bW 50\% of the time.

The ATLAS Collaboration published a search for single production of Y and
T quarks decaying into bW using 8\TeV proton-proton collisions~\cite{ATLAS-singleWb}.
The analysis presented here is the first such search using 13\TeV proton-proton data,
and sets the most stringent limits to date on the production cross section
for a single Y or T quark.
The search is carried out based on events containing one electron or muon, at least one b-tagged jet with large transverse momentum (\pt), at least one jet in the
forward region of the detector, and missing transverse momentum.

\section{CMS detector and event samples}
\label{sec:data}
The essential feature of the CMS detector is the superconducting
solenoid, 6\unit{m} in diameter and 13\unit{m} in length, which provides an axial
magnetic field of 3.8\unit{T}.
Within the solenoid volume a multi-layered silicon pixel and strip tracker
is used to measure the trajectories of charged particles with pseudorapidity
$\abs{\eta} < 2.5$. Outside of the tracker system, an electromagnetic calorimeter
(ECAL) made of lead tungstate crystals and a hadron calorimeter (HCAL)
made of brass and scintillators cover the region $\abs{\eta}<3.0$.
The region $3.0<\abs{\eta}<5.0$ is covered by the forward hadronic calorimeter, which is made primarily of steel and quartz fibers.
Muons are measured in gas-ionization detectors embedded in the steel flux-return yoke of the solenoid, and covering
the region $\abs{\eta} < 2.4$.
A more detailed description of the CMS detector, together with a definition of the coordinate system used and the relevant kinematic variables,
can be found in Ref.~\cite{CMSdet}.

The data used for this analysis were recorded during the 2015 data taking
period in proton-proton collisions at a center-of-mass energy of 13\TeV,
corresponding to an integrated luminosity of 2.3\fbinv.
The electron data sample was collected using a trigger that required at least one isolated electron with $\abs{\eta} <$ 2.5 and \pt $>$ 27\GeV.
The muon data sample was collected using a trigger that required at least one isolated muon with $\abs{\eta} <$ 2.1 and \pt $>$ 20\GeV.

The VLQ signal efficiencies and background contributions are estimated using Monte Carlo (MC) samples.
They are validated using background enriched data samples.
The $\ttbar$+jets, t- and $\PQt \PW$-channel single top-quark production and the
$\PW \PW$ processes are simulated using {\POWHEG} v2~\cite{powheg1,powheg2,powheg3}.
Single top quark production via s-channel and the $\PW \PZ$ process
are simulated with \MCatNLO v2~\cite{amcatnlo}. Inclusive boson production
($\PW$+jets and $\PZ$+jets) is simulated with {\MADGRAPH}
v5~\cite{madgraph}. {\PYTHIA} 8.212~\cite{pythia, pythia8} is used for
parton shower development and hadronization and to simulate
QCD multijet events.

The VLQ processes considered in this paper are generated using the tree-level MC event generator {\MADGRAPH} v5 for VLQ masses in the range from 0.70 to 1.80\TeV,
in steps of 100\GeV. The VLQ width is set to 10\GeV for all masses. The NNPDF3.0~\cite{nnpdf} parton distribution functions (PDFs) are used for both signal
and SM MC processes to model the momentum distribution of the colliding partons inside the protons.

The cross sections used to normalize the SM processes are calculated to
next-to-leading order (NLO) or to next-to-next-to-leading order (NNLO),
where the latter is available~\cite{sm_ww, sm_ttbar, sm_top}. For the signal,
the NLO cross sections are taken from Refs.~\cite{single_cross_sec1, single_cross_sec2}.
For the $\ttbar$+jets, $\PQt \PW$-channel single top-quark, and $\PW \PW$ SM processes, NNLO cross sections are used,
while NLO cross sections are applied to the remaining processes.

All generated events are processed through the CMS detector simulation
based on \GEANTfour~\cite{geant4}. Additional minimum bias events,
generated with \PYTHIA 8.212, are superimposed on the hard-scattering events
to simulate multiple proton-proton interactions (pileup) within the neighboring bunch
crossings.
The simulated events are weighted to reproduce the distribution of the number of
pileup interactions, 20 on average, observed in data.

\section{Event reconstruction}
\label{sec:eventReco}
All physics objects in the event are reconstructed using a particle-flow
(PF) algorithm~\cite{particleflow, particleflow2}, which
uses information from all subsystems to reconstruct photons, electrons,
muons, and charged and neutral hadrons.
Charged particle tracks are used to reconstruct the interaction vertices.
The vertex with the highest sum of squared \pt of all associated tracks
is taken as the primary vertex of the hard collision.
Filters are applied to reject events where electronic noise
or proton-beam backgrounds mimic energy deposits in the detector.

Electron candidates are reconstructed by combining the tracking information
with energy deposits in the ECAL in the range $\abs{\eta} < 2.5$
(excluding the range $1.4442 < \abs{\eta} < 1.566$, which is a transition region between
endcap and barrel calorimeters).
Tight identification criteria are applied to select well-reconstructed
electron candidates.
Candidates are identified~\cite{CMS:electrons} using
information on the shower-shape, the track quality and the spatial match
between the track and the electromagnetic cluster, the fraction of total
cluster energy in the HCAL, and the resulting level of
activity in the surrounding tracker and calorimeter regions.
The energy resolution for electrons with $\pt > 40\GeV$, measured using $\PZ \to \Pe \Pe$
decays, is on average 1.7\% in the ECAL central region of the detector~\cite{CMS:electrons}.

Muon candidates are identified using track segments reconstructed separately from hits in the silicon tracking system and in the muon system.
To identify muon candidates, the track segments must be consistent with muons originating from the primary vertex and satisfying
tight identification requirements. The matching of the muon and silicon track segments results in a relative \pt resolution
of 1.3$-$2.0\% in the central region of the detector for muons with $20 < \pt < 100\GeV$, and for muons with \pt up to 1\TeV the resolution is
10\% or better~\cite{CMS:muons}.

Lepton (electron or muon) reconstruction and trigger efficiencies are evaluated
as a function of \pt and $\abs{\eta}$ in both data and simulation, using a ``tag-and-probe''
method~\cite{TagAndProbe} with recorded and simulated samples of dileptonic Z events.

An isolation variable is employed to suppress leptons
originating from QCD processes. We define a relative isolation as the sum of the \pt
of particle tracks found in the tracker and energy deposits found in the calorimeters within
a cone $\Delta R = \sqrt{\smash[b]{(\Delta\eta)^{2} + (\Delta\phi)^{2}}} = $ 0.3 (0.4) around
the trajectory of the electron (muon), divided by the lepton \pt.
Relative isolation is corrected for the effects of pileup, and is required to be less than 0.15 for muons, and less than 0.4 (0.6) for electrons in the barrel (endcap) region.

Particles reconstructed by the PF algorithm are clustered into jets by
using the direction of each particle at the interaction vertex. Charged hadrons
found by the PF algorithm that are associated with pileup vertices are not
considered.
Particles that are identified as isolated leptons are removed from
the jet clustering procedure.
Jets are reconstructed with the
anti-$\kt$ algorithm~\cite{Cacciari:dispelling, antikt} with a distance parameter of 0.4.
An event-by-event jet-area-based correction~\cite{fastjet1,
fastjet2} is applied to remove, on a statistical basis, neutral pileup contribution
that is not already removed by the charged-hadron subtraction
procedure described above. Jet energy corrections are applied to each jet, as a function of
\pt and $\eta$, to correct for the calorimeter response~\cite{Chatrchyan:2011ds}.

The missing transverse momentum is defined as
the negative vector sum of the transverse momenta of all the particles
found by the PF algorithm, and its magnitude is referred to as $\MET$.
The decay of a heavy quark into a leptonically decaying $\PW$ boson and a b quark is expected to exhibit
genuine missing transverse momentum because of the undetected neutrino from the $\PW$ decay.
A missing transverse momentum threshold is applied to the selected
events, and the missing transverse momentum vector is used in the mass
reconstruction.

To identify jets originating from a b quark (b-tagged jets), the combined
secondary vertex (CSV) algorithm is used~\cite{CMS-PAS-BTV-15-001,BTV_12_paper}.
This tagging algorithm combines variables that can distinguish b quark jets from those originating
from light flavors, such as information on track impact parameter significance and
secondary vertex properties. The variables are combined using a
likelihood ratio technique to compute a b tagging discriminator.
We use the CSV medium operating point~\cite{CMS-PAS-BTV-15-001}, which achieves
a b tagging efficiency of approximately  $70\%$ and a mistag rate
of $1\%$. Data-to-Simulation efficiency and mistag rate scale
factors account for the small differences
observed between data and simulation.
We use these scale factors as a function of jet \pt and
$\eta$~\cite{CMS-PAS-BTV-15-001} to correct simulated events.

\section{Event selection and search strategy}
\label{sec:eventSele}

The signal event selection requires exactly one lepton with $\pt > 40\GeV$ and
$\abs{\eta} < 2.1$. Events with additional leptons having $ \pt > 10\GeV$ and $\abs{\eta} < 2.5$ and passing relatively loose isolation and
identification requirements are rejected to suppress dileptonic events.

Events are required to have at least two jets, one in the central and one
in the forward region of the detector. The central jet is required to have
$\pt  > 200\GeV$ and $\abs{\eta} < 2.4$ and be b-tagged.
When there is more than one central jet satisfying the above criteria, the leading central jet is used to reconstruct the mass of the VLQ.
The forward jet ($2.4 < \abs{\eta} < 5.0$) must have $\pt  > 30\GeV$.

In the decay of a singly produced VLQ, the b quark and the W boson tend to be produced with
the transverse momenta pointing in opposite directions.
Hence, the azimuthal angle between the central b jet and the lepton $\ell$ is
required to satisfy $\Delta \phi(\ell, \PQb) > 2$.
In addition, the lepton is required to be separated from any jets with $\pt > 40\GeV$ produced
in the event. When a hadronic jet is found  within $\Delta R(\ell, \text{jet}) <  1.5$,
the event is rejected. Since W boson originating from heavy VLQ decay has significant \pt,
events are required to have substantial $\MET$ ($> 50\GeV$) due to the undetected neutrino from the W boson decay.
The transverse mass, $M_{\mathrm{T}}$, formed by
the lepton and $\MET$ system is required to satisfy $M_{\mathrm{T}} < 130\GeV$ to
suppress $\ttbar$ dilepton events, which can mimic the
signal when one of the leptons escapes detection.

Finally, events are required to have $S_{\mathrm{T}} > 500\GeV$, where
$S_{\mathrm{T}}$ is defined as the scalar sum of the transverse momenta of the lepton, the leading central jet, and the missing transverse momentum.
This requirement reduces the signal efficiency by less than 10\% for the VLQ mass range considered in this paper.

The invariant  mass of the heavy quark candidate, $M_\text{inv}$, is reconstructed from  its decay products: the lepton, the leading central jet,
and the neutrino, where the $x$,$y$-components of the neutrino momentum are given
by the missing transverse momentum, while the $z$-component  is determined by constraining the invariant mass of
the lepton and neutrino to the W boson mass value. The solution with the smallest value is considered as the $z$-component. This method is used only when the solution
of the relevant quadratic equation is real, otherwise the $z$-component is set to zero.

The single VLQ production $\Y/\T \to \PQb \PW$ would result
in a peak in the $M_{\mathrm{inv}}$ distribution at the mass of the VLQ. The experimental mass resolution is 12$-$15\% and is independent of the VLQ mass.

\section{Background modeling}

\begin{figure}[thb]
\centering
\includegraphics[width=0.49\textwidth]{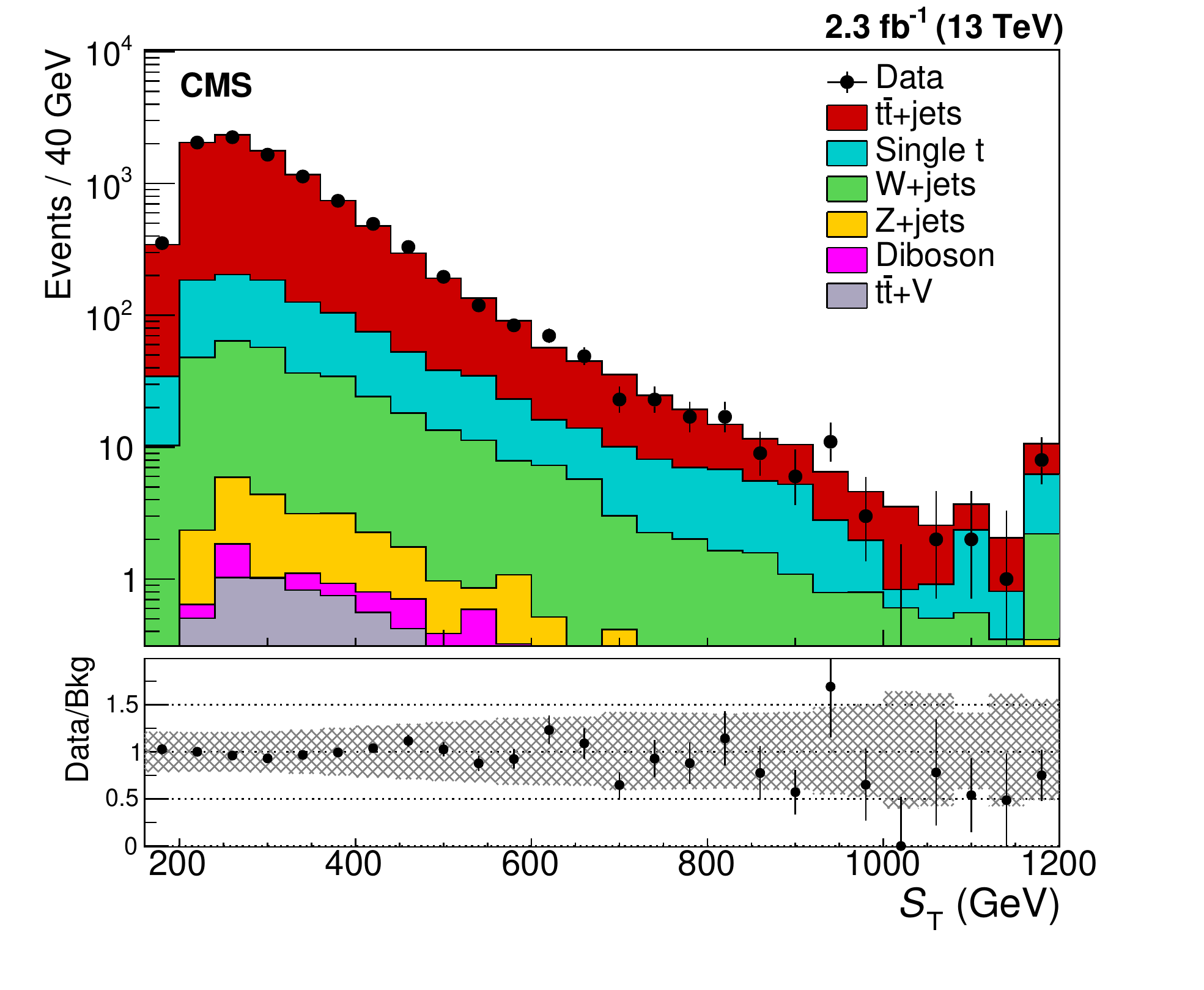}
\includegraphics[width=0.49\textwidth]{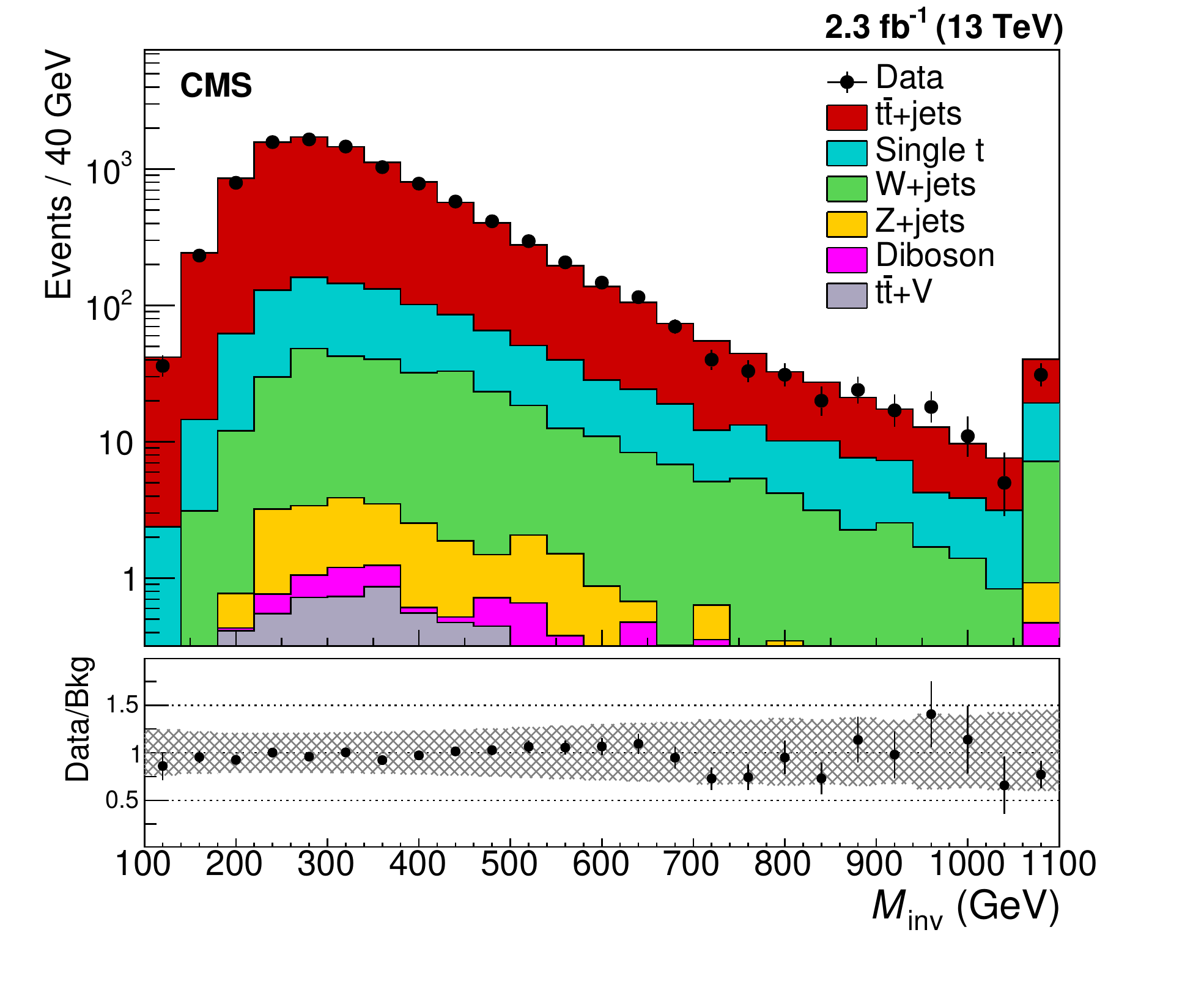}
\caption{Kinematic distributions in the $\ttbar$-enriched control sample: $S_\mathrm{T}$ (\cmsLeft) and $M_{\mathrm{inv}}$ (\cmsRight). The last bin includes overflow events.  The statistical and systematic uncertainties
are represented by the hatched band on the ratio plot.}
\label{fig:ttbar}
\end{figure}

\label{sec:bkgSel}
The dominant background processes in this search are the production of $\ttbar$ and $\PW$+jets events.
The modeling of these processes is validated by
studying background-enriched samples.

To verify the modeling of the $\ttbar$ process, we select
events with the lepton and \MET fulfilling the signal selection criteria, and
at least 2 b-tagged jets with the leading (sub-leading)
jet satisfying the requirement of $\pt > 70\, (30)\GeV$. We also remove the
$\Delta R(\ell, \text{jet})$, $\Delta \phi(\ell, \PQb)$ and forward jet requirements to enrich the sample with $\ttbar$ events.

The top quark \pt spectrum from the $\ttbar$ simulation is
known to be mismodeled and is reweighted using the empirical function described in Ref.~\cite{diff-ttbar}.
After this correction, the data points at large values of all relevant kinematic distributions are consistent within systematic uncertainties.
Distributions of $S_\mathrm{T}$ and the invariant mass of the bW system in the $\ttbar$ sample are shown in Fig.~\ref{fig:ttbar}.

\begin{figure}[thb]
\centering
\includegraphics[width=0.49\textwidth]{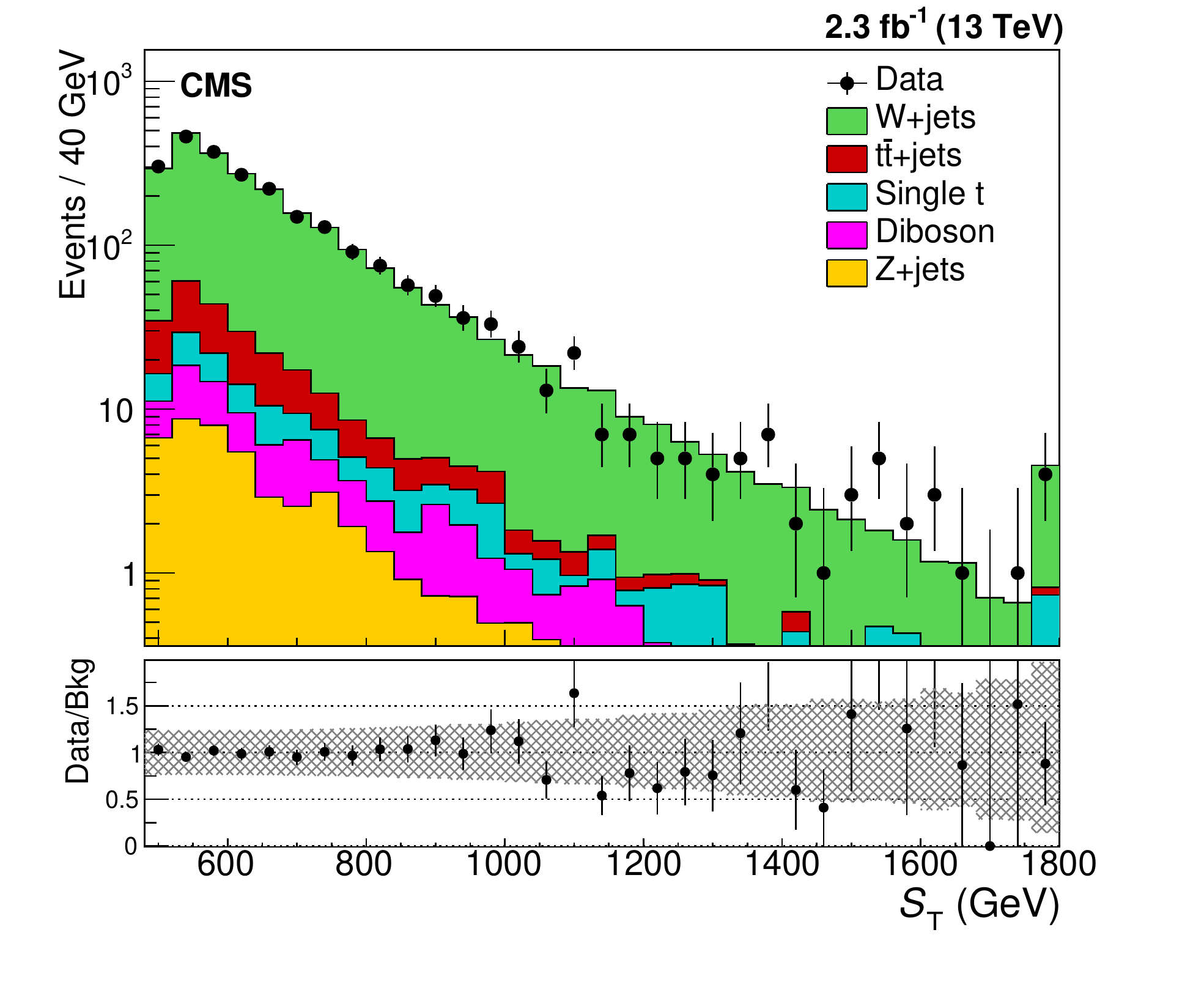}
\includegraphics[width=0.49\textwidth]{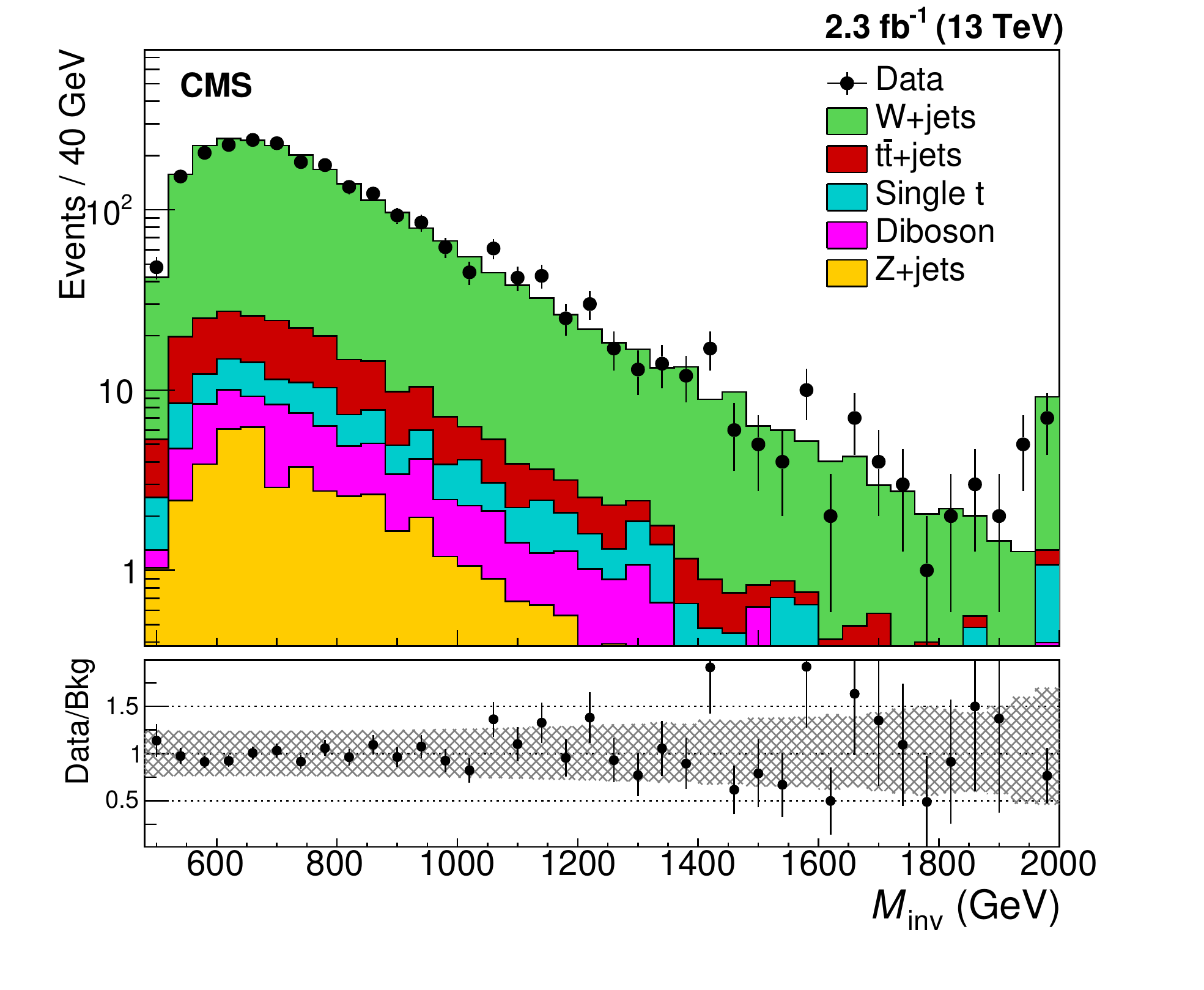}
\caption{Kinematic distributions in the $\PW$+jets-enriched control sample: $S_\mathrm{T}$ (\cmsLeft) and $M_{\text{inv}}$ (\cmsRight). The last bin includes overflow events. The statistical and systematic uncertainties are represented by the hatched band on the ratio plot.}
\label{fig:Wjets}
\end{figure}

The $\PW$+jets-enriched control sample requirement is identical to the signal event selection except that events with b-tagged jets are vetoed.
We observe that in the $\PW$+jets simulated sample, the number of events at large
jet \pt distributions is overestimated as compared with the distributions measured in data.
We derive a correction for the $\PW$+jets simulation as a function of the $H_\mathrm{T}$ variable, defined as the scalar
sum of the transverse momenta of all jets with $\pt  >  30\GeV$.
The data to simulation ratio of the $H_\mathrm{T}$ distribution is well described by a 2-parameter linear fit with a negative slope.
A correction to the modeling of the $\PW$+jets $H_\mathrm{T}$ spectrum is made using the results of the fit.
After the correction is applied, good agreement in the modeling of all kinematic variables is observed.
Distributions of $S_\mathrm{T}$ and the invariant mass of the bW system in the $\PW$+jets sample are shown in Fig.~\ref{fig:Wjets}.

\section{Systematic uncertainties}
\label{sec:uncertainties}
We divide the systematic uncertainties
into two categories: uncertainties that
impact only the rate of background and signal predictions, and
uncertainties that affect both the rate and the shape of the fitted $M_{\text{inv}}$ spectra.
The shape uncertainties affecting the $M_{\text{inv}}$ distribution are modeled by varying
the nuisance parameters that characterize the associated systematic effects up and down by one standard deviation.

The uncertainty in the integrated luminosity is 2.7\%~\cite{CMS_lumi}.
We assign the uncertainties for the normalization of the SM background processes as the uncertainties on corresponding CMS cross section measurements at 13\TeV, which are 5.6\% for $\ttbar$~\cite{cms_ttbar}, 14.7\% for single top quark~\cite{cms_top_paper}, and 9.2\% for $\PW$+jets~\cite{CMS-PAS-SMP-15-004}, where in the last case we also account separately for uncertainties in the W+heavy-flavor contributions~\cite{WplusC, WplusB_paper}.

To account for the MC mismodeling correction in the $\PW$+jets sample, we derive a two-sided uncertainty band using the $H_\mathrm{T}$ correction procedure. To account for the MC mismodeling correction in the $\ttbar$ sample, we derive a two-sided uncertainty band using the top \pt reweighting procedure. One side of the band is obtained by removing the correction, and the other side is obtained by applying the procedure twice.
The uncertainties due to these corrections increase with the rise of the top quark \pt and $H_\mathrm{T}$, which leads to the widening of the uncertainty band at large $S_\mathrm{T}$ and $M_{\text{inv}}$, as can be seen in Figs.~\ref{fig:ttbar} and~\ref{fig:Wjets}.

In addition, the reconstruction efficiency of forward jets has been observed to be
larger in the simulation than in the data. The efficiency as a function of $\eta$ is corrected to match the data using the $\PW$+jets-enriched sample with 0 b-tagged jets, and validated using the $\ttbar$-enriched sample with two b-tagged jets. An uniform rate uncertainty of ${\pm}15\%$ is assigned to cover the forward jet mismodeling in simulation.

Trigger and lepton identification efficiencies in simulation
are corrected as functions of lepton \pt and $\eta$ using decays of $\PZ$ bosons to leptons in data.
The associated uncertainty of about 2\% is the statistical uncertainty in the data.

The shape uncertainties include uncertainties in the jet energy scale, jet energy resolution,
$\PQb$ tagging efficiency,
pileup,
PDFs, as well as
factorization and renormalization scales.
These uncertainties are treated as uncorrelated.

The uncertainty related to the modeling of pileup
is evaluated by varying the inelastic cross section
by ${\pm}5\%$ relative to the nominal value of 69\unit{mb}~\cite{lumi}.
Uncertainties in renormalization and factorization scales are
taken into account by varying both scales simultaneously
up and down by a factor of two. Uncertainties arising from
the choice of PDFs are taken into account
according to the PDF4LHC procedure~\cite{pdf_lhc}.

The systematic uncertainties are summarized in Table~\ref{tab:SystTable}.

\begin{table*}[thb]
\renewcommand{\arraystretch}{1.1}
\centering
\topcaption{Summary of the systematic uncertainties associated with the simulated backgrounds and the signal events. The value quoted represents the expected change in the event yield in the signal region due to the systematic uncertainty.}
\label{tab:SystTable}
\begin{tabular}{lccccc} \hline
Source &   & $\PW$+jets &  $\ttbar$ & Single top & Signal    \\
\hline
Integrated luminosity & rate & 2.7\% & 2.7 \% & 2.7 \% & 2.7 \% \\
Jet energy scale & shape & 5\%  & 6\%  & 5\%  & 3\%  \\
Jet energy resolution & shape & 2\%  & 1\% & 1\% & 2\% \\
b tagging efficiency & shape & 3\%  & 5\% & 5\% & 5\% \\
Multiple interactions& shape & 1\% & 1\% & 1\%   & 1\%  \\
Lepton efficiency & rate & 2\% & 2\% & 2\%  & 2\% \\
Trigger efficiency & rate & 2\% & 2\% & 2\% & 2\% \\
Cross section & rate & 9.2\% & 5.6\% & 14.7\%& --- \\
Top quark \pt reweighting & shape & --- & 28\% & --- & --- \\
$\PW$+jets $H_\mathrm{T}$ reweighting & shape & 5.3\% & --- & --- & --- \\
Renormalization/factorization scales & shape & 14\% & 16\% & 16\% & --- \\
PDF & shape & 5.5\% & 2.3\% & 8.5\% & 6.7\% \\
Forward jet reweighting & rate & 15\% & 15\% & 15\% & 15\% \\ \hline
\end{tabular}

\end{table*}

\begin{table*}[btp]
\centering
\renewcommand{\arraystretch}{1.1}
\topcaption{Data, background, and possible signal pre-fit event yields corresponding to 2.3\fbinv of integrated luminosity.
The signal sample is the $M(\Y)=1.0\TeV$ mass point using the NLO cross~section~\cite{single_cross_sec2}.
The percentage in the signal column indicates the signal efficiency.
The background uncertainties include both the statistical and the systematic pre-fit components.}
\label{tab:srnumbers}
\resizebox{\textwidth}{!}{
\begin{tabular}{cccccccccc}
\hline
Channel &  $\PW$+jets  & $\ttbar$ &  Single t &  QCD &  $\PZ$+jets & Diboson & Y\,(1.0\TeV) &  Total bkg. & Data \\
\hline
Electron& $44 \pm 12$ & $28 \pm 11$ & $20 \pm 5$ & ${<}1$ & $1.5 \pm 1.5$ & $1.3 \pm 0.5$ & $54\,(1.3\%)$ & $\phantom{0}95 \pm 17$  & $78$ \\
Muon&     $52 \pm 14$ & $34 \pm 13$ & $27 \pm 6$ & ${<}1$ & ${<}1$ & $1.7 \pm 0.6$ & $60\,(1.4\%)$ & $115 \pm 20$ & $95$ \\ \hline
\end{tabular}
}

\end{table*}

\section{Limit calculation and results}
\label{sec:results}
\begin{figure}[htp]
\centering
\includegraphics[width=\cmsFigWidth]{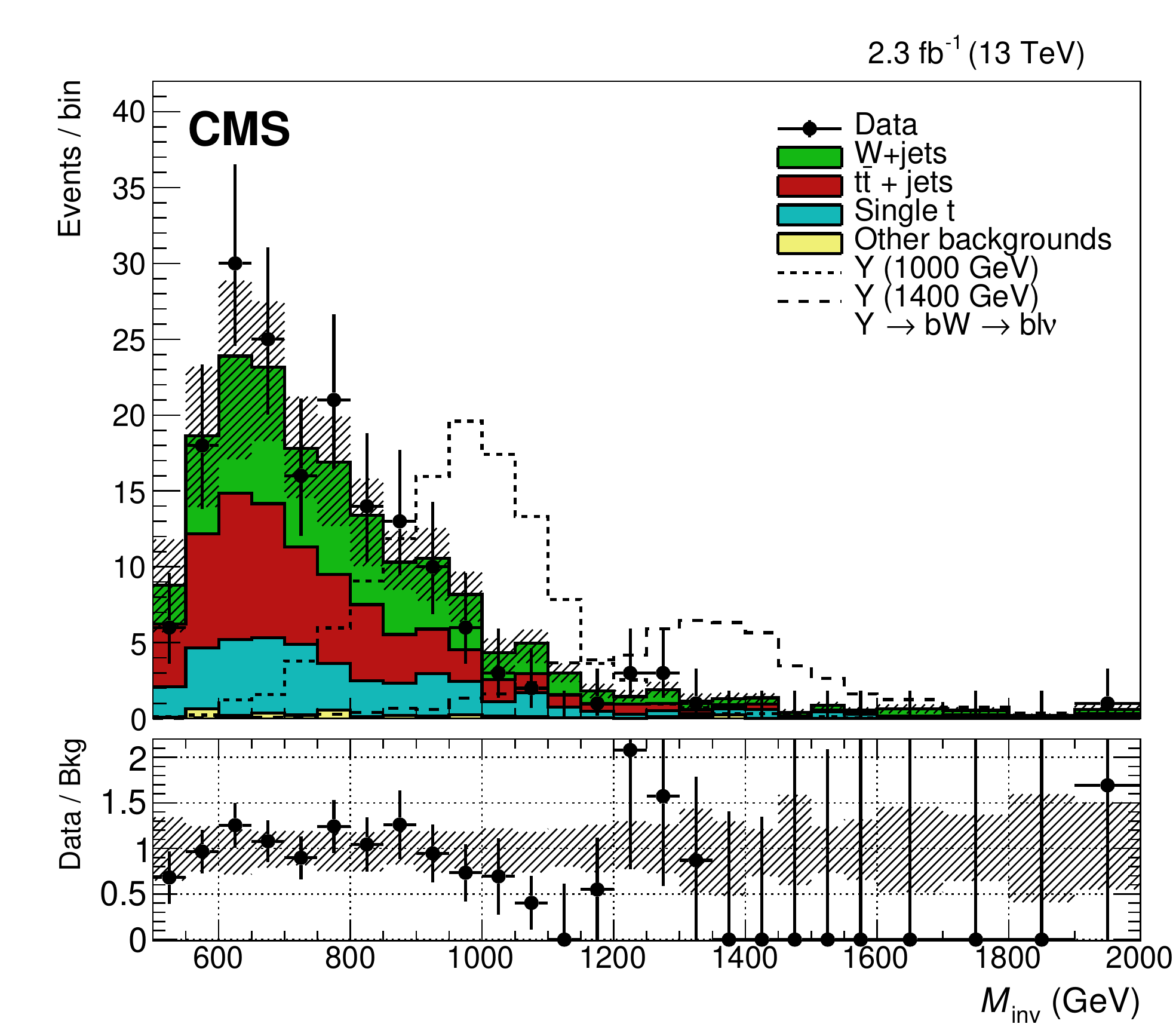}
\caption{The invariant mass $M_{\text{inv}}$ distribution of heavy quark candidates, reconstructed from their decay products: the lepton, the leading central jet, and
the neutrino. The distribution is obtained after the fit, assuming the background-only hypothesis.
The dashed histograms show the event distributions expected for a Y quark with masses  of 1.0\TeV and 1.4\TeV, coupling of 0.5 and ${\mathcal{B}}(\Y \to \PQb \PW) = 100\%$.
The statistical and systematic uncertainties are represented by the hatched band on the ratio plot. In the last bin the data overflow event is an electron channel event with a mass of 2.22\TeV. }
\label{fig:mass_post_fit}
\end{figure}

Good agreement between the event yields in the data and in the SM prediction is observed within uncertainties, as shown in Table~\ref{tab:srnumbers}.
The sum of the SM backgrounds and a hypothesized signal for the combined electron and muon channels is fitted to the observed spectrum of $M_{\text{inv}}$.
The fit uses a binned likelihood method, where the binning of the distributions is chosen in such a way that the statistical uncertainty in the MC estimation of total background per bin is always less than 20\%.
Contributions from the SM processes are allowed to
float independently within their systematic uncertainties,
using log-normal priors~\cite{log_prior_ref1,log_prior_ref2}. The nuisance parameters describing the shape
uncertainties are constrained using Gaussian priors.
The shapes of the $M_{\text{inv}}$ distributions for backgrounds and signal
are parametrized and varied according to the nuisance parameters.
The post-fit $M_{\text{inv}}$  distribution, with the shape and background
normalizations corresponding to the maximum likelihood values, is presented
in Fig.~\ref{fig:mass_post_fit}. All corrections derived from the background-enriched regions are propagated to the signal region and appropriate systematic
uncertainties are assigned.

\begin{figure}[htp]
\centering
\includegraphics[width=\cmsFigWidth]{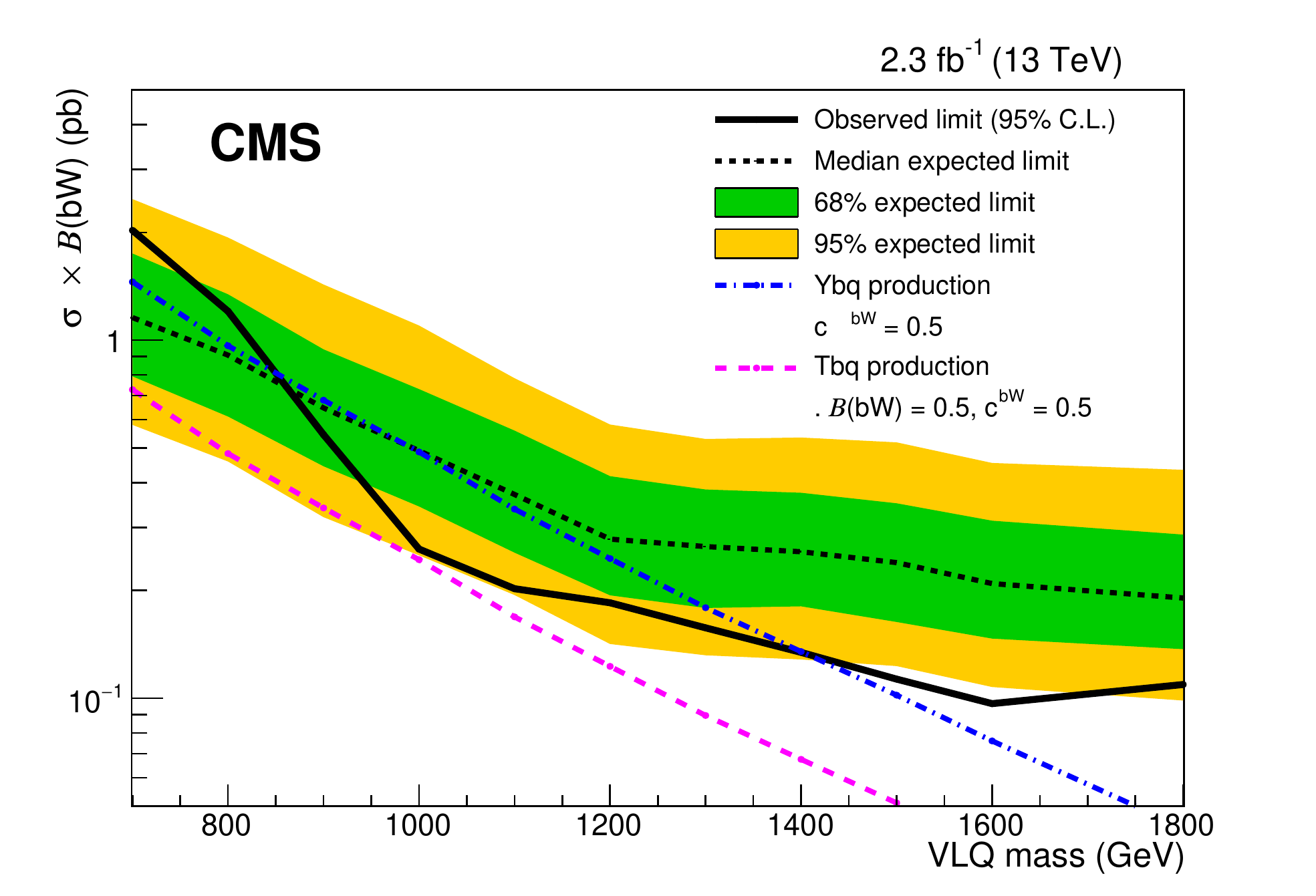}
\caption{Expected and observed limits on the single VLQ production ($\Pp \Pp\to \Y \PQb \PQq$ and $\Pp \Pp \to \T \PQb \PQq$) cross section together with the one and two standard deviation uncertainty bands. }
\label{fig:limit}
\end{figure}

Upper limits  at 95\% confidence level (CL) on the production cross section of the $\Y/ \T \to \PQb \PW$
process are computed using a Bayesian approach~\cite{Theta_web}, where the likelihood
is marginalized with respect to the nuisance parameters representing systematic
uncertainties.
The expected limit is calculated by resampling the data from the background distribution.
The 95\% CL expected and observed upper limits are listed in Table~\ref{tab:upperlimits} and shown in Fig.~\ref{fig:limit}.
The observed limits at high VLQ mass reflect a 2$\sigma$ deficit of events above 1.0 TeV in the $M_{\text{inv}}$ distribution.
The limits are derived assuming a narrow width for the VLQ.
The VLQ width is proportional to the square of the coupling, and is negligible compared to the experimental resolution for couplings
below 0.5, for the range of VLQ masses considered in this paper.
In the framework of the model considered, Y quarks with a coupling of 0.5 and ${\mathcal{B}}(\Y \to \PQb \PW) = 100\%$ are excluded in the mass range from 0.85 to
1.40\TeV. This result may be compared with the expected region of excluded masses, which extends up to 1.0\TeV.
In the case of T quarks with a coupling of 0.5,
the theoretical cross  section, the selection efficiency and the $M_{\text{inv}}$ distribution are the same as those for the production and decay of Y quarks,
but the expected decay branching fraction $\mathcal{B}(\T \to \PQb \PW)$ is 50\%, only half that expected for ${\mathcal{B}}(\Y \to \PQb \PW)$. Thus mass
exclusion limits similar to those achieved for the Y quark would only be obtained for ${\mathcal{B}}(\T \to \PQb \PW) = 100\%$.

\begin{table}[ht]
\renewcommand{\arraystretch}{1.2}
\centering
\topcaption{The 95\% CL expected and observed upper limits (UL)  on the single VLQ production cross section, assuming ${\mathcal{B}}( \mathrm{VLQ} \to \PQb \PW )=100\%$.}
\label{tab:upperlimits}
\begin{tabular}{ccc} \hline
VLQ mass & Expected UL  & Observed UL \\ [-0.6ex]
(\TeVns{}) & (pb) & (pb) \\
\hline
0.70 & $1.16^{+0.68}_{-0.37}$   & 2.03 \\
0.80 & $0.91^{+0.43}_{-0.30}$ & 1.20 \\
0.90 & $0.65^{+0.29}_{-0.21}$  & 0.54 \\
1.0\phantom{0} & $0.49^{+0.24}_{-0.15}$ & 0.26 \\
1.10 & $0.37^{+0.19}_{-0.12}$ & 0.20 \\
1.20 & $0.28^{+0.14}_{-0.09}$ & 0.18 \\
1.30 & $0.27^{+0.11}_{-0.09}$ & 0.16 \\
1.40 & $0.26^{+0.11}_{-0.08}$ & 0.13 \\
1.50 & $0.24^{+0.11}_{-0.08}$ & 0.11 \\
1.60 & $0.21^{+0.11}_{-0.06}$ & 0.10 \\
1.70 & $0.20^{+0.10}_{-0.06}$ & 0.10 \\
1.80 & $0.19^{+0.09}_{-0.05}$ & 0.11 \\ \hline
\end{tabular}
\end{table}

\section{Summary}
\label{sec:summary}
A search has been performed for single production of a vector-like quark decaying into a b quark and a W boson in the electron/muon + jets channels. The mass of the vector-like quark is reconstructed by forming the invariant mass of the leading b-tagged jet, electron or muon, and missing transverse momentum in the event, and a fit to the invariant mass spectrum is performed. No evidence of an excess due to new physics is observed. Upper limits at 95\% CL are set on the cross sections for single production of vector-like Y and T quarks in the mass range from 0.70 to 1.80\TeV. In the framework of the model considered, Y quarks with a coupling of 0.5 and ${\mathcal{B}}(\Y \to \PQb \PW) = 100\%$ are excluded in the mass range from 0.85 to 1.40\TeV. This result may be compared with the expected region of excluded masses, which extends up to 1.0\TeV. These results represent the most stringent limits to date on the single production of a vector-like Y quark.  In the case of T quarks with a coupling of 0.5,
the theoretical cross  section, the selection efficiency and the $M_{\text{inv}}$ distribution are the same as those for the production and decay of Y quarks,
but the expected decay branching fraction ${\mathcal{B}}(\T \to \PQb \PW)$ is 50\%, only half that expected for ${\mathcal{B}}(\Y \to \PQb \PW)$. Thus mass exclusion limits similar to those achieved for the Y quark would only be obtained for ${\mathcal{B}} ( \T \to \PQb \PW) = 100\%$.

\begin{acknowledgments}
We congratulate our colleagues in the CERN accelerator departments for the excellent performance of the LHC and thank the technical and administrative staffs at CERN and at other CMS institutes for their contributions to the success of the CMS effort. In addition, we gratefully acknowledge the computing centers and personnel of the Worldwide LHC Computing Grid for delivering so effectively the computing infrastructure essential to our analyses. Finally, we acknowledge the enduring support for the construction and operation of the LHC and the CMS detector provided by the following funding agencies: BMWFW and FWF (Austria); FNRS and FWO (Belgium); CNPq, CAPES, FAPERJ, and FAPESP (Brazil); MES (Bulgaria); CERN; CAS, MoST, and NSFC (China); COLCIENCIAS (Colombia); MSES and CSF (Croatia); RPF (Cyprus); SENESCYT (Ecuador); MoER, ERC IUT, and ERDF (Estonia); Academy of Finland, MEC, and HIP (Finland); CEA and CNRS/IN2P3 (France); BMBF, DFG, and HGF (Germany); GSRT (Greece); OTKA and NIH (Hungary); DAE and DST (India); IPM (Iran); SFI (Ireland); INFN (Italy); MSIP and NRF (Republic of Korea); LAS (Lithuania); MOE and UM (Malaysia); BUAP, CINVESTAV, CONACYT, LNS, SEP, and UASLP-FAI (Mexico); MBIE (New Zealand); PAEC (Pakistan); MSHE and NSC (Poland); FCT (Portugal); JINR (Dubna); MON, RosAtom, RAS, RFBR and RAEP (Russia); MESTD (Serbia); SEIDI, CPAN, PCTI and FEDER (Spain); Swiss Funding Agencies (Switzerland); MST (Taipei); ThEPCenter, IPST, STAR, and NSTDA (Thailand); TUBITAK and TAEK (Turkey); NASU and SFFR (Ukraine); STFC (United Kingdom); DOE and NSF (USA).

\hyphenation{Rachada-pisek} Individuals have received support from the Marie-Curie program and the European Research Council and EPLANET (European Union); the Leventis Foundation; the A. P. Sloan Foundation; the Alexander von Humboldt Foundation; the Belgian Federal Science Policy Office; the Fonds pour la Formation \`a la Recherche dans l'Industrie et dans l'Agriculture (FRIA-Belgium); the Agentschap voor Innovatie door Wetenschap en Technologie (IWT-Belgium); the Ministry of Education, Youth and Sports (MEYS) of the Czech Republic; the Council of Science and Industrial Research, India; the HOMING PLUS program of the Foundation for Polish Science, cofinanced from European Union, Regional Development Fund, the Mobility Plus program of the Ministry of Science and Higher Education, the National Science Center (Poland), contracts Harmonia 2014/14/M/ST2/00428, Opus 2014/13/B/ST2/02543, 2014/15/B/ST2/03998, and 2015/19/B/ST2/02861, Sonata-bis 2012/07/E/ST2/01406; the National Priorities Research Program by Qatar National Research Fund; the Programa Clar\'in-COFUND del Principado de Asturias; the Thalis and Aristeia programs cofinanced by EU-ESF and the Greek NSRF; the Rachadapisek Sompot Fund for Postdoctoral Fellowship, Chulalongkorn University and the Chulalongkorn Academic into Its 2nd Century Project Advancement Project (Thailand); and the Welch Foundation, contract C-1845.
\end{acknowledgments}

\clearpage
\bibliography{auto_generated}

\providecommand{\href}[2]{#2}\begingroup\raggedright\begin{thebibliography}{10}%
\makeatletter
\providecommand{\hrefCMSnoop }[0]{\@secondoftwo}%
\makeatother
\providecommand{\doi}{\texttt{doi:}\begingroup \urlstyle{tt}\Url}

\bibitem{Atlas_Higgs}
\hrefCMSnoop {}{{ATLAS Collaboration}, ``{Observation of a new particle in the
  search for the Standard Model Higgs boson with the ATLAS detector at the
  LHC}'',} \textit{ Phys. Lett. B} \textbf{ 716} (2012) 1,
  \href{http://dx.doi.org/10.1016/j.physletb.2012.08.020}{\doi{10.1016/j.physletb.2012.08.020}},
  \href{http://www.arXiv.org/abs/1207.7214}{\texttt{arXiv:1207.7214}}.

\bibitem{CMS_Higgs}
\hrefCMSnoop {}{{CMS Collaboration}, ``{Observation of a new boson at a mass of
  125\GeV with the CMS experiment at the LHC}'',} \textit{ Phys. Lett. B}
  \textbf{ 716} (2012) 30,
  \href{http://dx.doi.org/10.1016/j.physletb.2012.08.021}{\doi{10.1016/j.physletb.2012.08.021}},
  \href{http://www.arXiv.org/abs/1207.7235}{\texttt{arXiv:1207.7235}}.

\bibitem{Higgs_long}
\hrefCMSnoop {}{{CMS Collaboration}, ``{Observation of a new boson with mass
  near 125\GeV in pp collisions at $\sqrt{s} =$ 7 and 8\TeV}'',} \textit{ JHEP}
  \textbf{ 06} (2013) 081,
  \href{http://dx.doi.org/10.1007/JHEP06(2013)081}{\doi{10.1007/JHEP06(2013)081}},
  \href{http://www.arXiv.org/abs/1303.4571}{\texttt{arXiv:1303.4571}}.

\bibitem{Lodone:2008yy}
\hrefCMSnoop {}{P.~Lodone, ``{Vector-like quarks in a 'composite' Higgs
  model}'',} \textit{ JHEP} \textbf{ 12} (2008) 029,
  \href{http://dx.doi.org/10.1088/1126-6708/2008/12/029}{\doi{10.1088/1126-6708/2008/12/029}},
\href{http://www.arXiv.org/abs/0806.1472}{\texttt{arXiv:0806.1472}}.

\bibitem{Perelstein:2003wd}
\hrefCMSnoop {}{M.~Perelstein, M.~E. Peskin, and A.~Pierce, ``{Top quarks and
  electroweak symmetry breaking in little Higgs models}'',} \textit{ Phys. Rev.
  D} \textbf{ 69} (2004) 075002,
  \href{http://dx.doi.org/10.1103/PhysRevD.69.075002}{\doi{10.1103/PhysRevD.69.075002}},
\href{http://www.arXiv.org/abs/hep-ph/0310039}{\texttt{arXiv:hep-ph/0310039}}.

\bibitem{Contino:2006nn}
\hrefCMSnoop {}{R.~Contino, T.~Kramer, M.~Son, and R.~Sundrum,
  ``{Warped/composite phenomenology simplified}'',} \textit{ JHEP} \textbf{ 05}
  (2007) 074,
  \href{http://dx.doi.org/10.1088/1126-6708/2007/05/074}{\doi{10.1088/1126-6708/2007/05/074}},
\href{http://www.arXiv.org/abs/hep-ph/0612180}{\texttt{arXiv:hep-ph/0612180}}.

\bibitem{TPrime}
\hrefCMSnoop {}{{CMS Collaboration}, ``{Search for vectorlike charge 2/3 T
  quarks in proton-proton collisions at $\sqrt{s} = 8\TeV$}'',} \textit{ Phys.
  Rev. D} \textbf{ 93} (2016) 012003,
  \href{http://dx.doi.org/10.1103/PhysRevD.93.012003}{\doi{10.1103/PhysRevD.93.012003}},
  \href{http://www.arXiv.org/abs/1509.04177}{\texttt{arXiv:1509.04177}}.

\bibitem{BPrime}
\hrefCMSnoop {}{{CMS Collaboration}, ``{Search for pair-produced vectorlike B
  quarks in proton-proton collisions at $\sqrt{s} = 8\TeV$}'',} \textit{ Phys.
  Rev. D} \textbf{ 93} (2016) 112009,
  \href{http://dx.doi.org/10.1103/PhysRevD.93.112009}{\doi{10.1103/PhysRevD.93.112009}},
  \href{http://www.arXiv.org/abs/1507.07129}{\texttt{arXiv:1507.07129}}.

\bibitem{ATLAS-TPrime}
\hrefCMSnoop {}{{ATLAS Collaboration}, ``{Search for production of vector-like
  quark pairs and of four top quarks in the lepton-plus-jets final state in pp
  collisions at $\sqrt{s} = 8\TeV$ with the ATLAS detector}'',} \textit{ JHEP}
  \textbf{ 08} (2015) 105,
  \href{http://dx.doi.org/10.1007/JHEP08(2015)105}{\doi{10.1007/JHEP08(2015)105}},
  \href{http://www.arXiv.org/abs/1505.04306}{\texttt{arXiv:1505.04306}}.

\bibitem{ATLAS-BPrime}
\hrefCMSnoop {}{{ATLAS Collaboration}, ``{Search for vectorlike B quarks in
  events with one isolated lepton, missing transverse momentum and jets at
  $\sqrt{s} = 8\TeV$ with the ATLAS detector}'',} \textit{ Phys. Rev. D}
  \textbf{ 91} (2015) 112011,
  \href{http://dx.doi.org/10.1103/PhysRevD.91.112011}{\doi{10.1103/PhysRevD.91.112011}},
  \href{http://www.arXiv.org/abs/1503.05425}{\texttt{arXiv:1503.05425}}.

\bibitem{DeSimone2013}
\hrefCMSnoop {}{A.~De~Simone, O.~Matsedonskyi, R.~Rattazzi, and A.~Wulzer, ``A
  first top partner hunter's guide'',} \textit{ JHEP} \textbf{ 04} (2013) 1304,
  \href{http://dx.doi.org/10.1007/JHEP04(2013)004}{\doi{10.1007/JHEP04(2013)004}},
  \href{http://www.arXiv.org/abs/1211.5663}{\texttt{arXiv:1211.5663}}.

\bibitem{handbook}
\hrefCMSnoop {}{J.~A. Aguilar-Saavedra, R.~Benbrik, S.~Heinemeyer, and
  M.~Perez-Victoria, ``A handbook of vectorlike quarks: mixing and single
  production'',} \textit{ Phys. Rev. D} \textbf{ 88} (2013) 094010,
  \href{http://dx.doi.org/10.1103/PhysRevD.88.094010}{\doi{10.1103/PhysRevD.88.094010}},
  \href{http://www.arXiv.org/abs/1306.0572}{\texttt{arXiv:1306.0572}}.

\bibitem{CMS_single_Ht_lep}
\hrefCMSnoop {}{{CMS Collaboration}, ``Search for single production of a heavy
  vector-like {T} quark decaying to a {H}iggs boson and a top quark with a
  lepton and jets in the final state'',} \textit{ Phys. Lett. B} \textbf{ 771}
  (2017) 80,
  \href{http://dx.doi.org/10.1016/j.physletb.2017.05.019}{\doi{10.1016/j.physletb.2017.05.019}},
  \href{http://www.arXiv.org/abs/1612.00999}{\texttt{arXiv:1612.00999}}.

\bibitem{CMS_single_Ht_had}
\hrefCMSnoop {}{{CMS Collaboration}, ``Search for electroweak production of a
  vector-like quark decaying to a top quark and a {H}iggs boson using boosted
  topologies in fully hadronic final states'',} \textit{ JHEP} \textbf{ 04}
  (2017) 136,
  \href{http://dx.doi.org/10.1007/JHEP04(2017)136}{\doi{10.1007/JHEP04(2017)136}},
  \href{http://www.arXiv.org/abs/1612.05336}{\texttt{arXiv:1612.05336}}.

\bibitem{CMS_single_Hz}
\hrefCMSnoop {}{{CMS Collaboration}, ``Search for single production of
  vector-like quarks decaying to a {Z} boson and a top or a bottom quark in
  proton-proton collisions at $\sqrt{s} = 13\TeV$'',} \textit{ JHEP} \textbf{
  05} (2017) 029,
  \href{http://dx.doi.org/10.1007/JHEP05(2017)029}{\doi{10.1007/JHEP05(2017)029}},
  \href{http://www.arXiv.org/abs/1701.07409}{\texttt{arXiv:1701.07409}}.

\bibitem{ATLAS-singleWb}
\hrefCMSnoop {}{{ATLAS Collaboration}, ``{Search for single production of
  vector-like quarks decaying into $Wb$ in $pp$ collisions at $\sqrt{s} =
  8\TeV$ with the ATLAS detector}'',} \textit{ Phys. J. C} \textbf{ 76} (2016)
  \href{http://dx.doi.org/10.1140/epjc/s10052-016-4281-8}{\doi{10.1140/epjc/s10052-016-4281-8}},
  \href{http://www.arXiv.org/abs/1602.05606}{\texttt{arXiv:1602.05606}}.

\bibitem{CMSdet}
\hrefCMSnoop {}{{CMS Collaboration}, ``{The CMS experiment at the CERN LHC}'',}
  \textit{ JINST} \textbf{ 3} (2008) S08004,
\href{http://dx.doi.org/10.1088/1748-0221/3/08/S08004}{\doi{10.1088/1748-0221/3/08/S08004}}.

\bibitem{powheg1}
\hrefCMSnoop {}{{P. Nason}, ``{A new method for combining NLO QCD with shower
  Monte Carlo algorithms}'',} \textit{ JHEP} \textbf{ 11} (2004) 040,
  \href{http://dx.doi.org/10.1088/1126-6708/2004/11/040}{\doi{10.1088/1126-6708/2004/11/040}},
  \href{http://www.arXiv.org/abs/hep-ph/0409146}{\texttt{arXiv:hep-ph/0409146}}.

\bibitem{powheg2}
\hrefCMSnoop {}{S.~Frixione, P.~Nason, and C.~Oleari, ``{Matching NLO QCD
  computations with parton shower simulations: the POWHEG method}'',} \textit{
  JHEP} \textbf{ 11} (2007) 070,
  \href{http://dx.doi.org/10.1088/1126-6708/2007/11/070}{\doi{10.1088/1126-6708/2007/11/070}},
  \href{http://www.arXiv.org/abs/0709.2092}{\texttt{arXiv:0709.2092}}.

\bibitem{powheg3}
\hrefCMSnoop {}{S.~Alioli, P.~Nason, C.~Oleari, and E.~Re, ``{A general
  framework for implementing NLO calculations in shower Monte Carlo programs:
  the POWHEG BOX}'',} \textit{ JHEP} \textbf{ 06} (2010) 043,
  \href{http://dx.doi.org/10.1007/JHEP06(2010)043}{\doi{10.1007/JHEP06(2010)043}},
  \href{http://www.arXiv.org/abs/1002.2581}{\texttt{arXiv:1002.2581}}.

\bibitem{amcatnlo}
J.~Alwall\hrefCMSnoop {}{ {et~al.}, ``{The automated computation of tree-level
  and next-to-leading order differential cross sections, and their matching to
  parton shower simulations}'',} \textit{ JHEP} \textbf{ 07} (2014) 079,
  \href{http://dx.doi.org/10.1007/JHEP07(2014)079}{\doi{10.1007/JHEP07(2014)079}},
  \href{http://www.arXiv.org/abs/1405.0301}{\texttt{arXiv:1405.0301}}.

\bibitem{madgraph}
J.~Alwall\hrefCMSnoop {}{ {et~al.}, ``{MadGraph 5: going beyond}'',} \textit{
  JHEP} \textbf{ 06} (2011) 128,
  \href{http://dx.doi.org/10.1007/JHEP06(2011)128}{\doi{10.1007/JHEP06(2011)128}},
  \href{http://www.arXiv.org/abs/1106.0522}{\texttt{arXiv:1106.0522}}.

\bibitem{pythia}
\hrefCMSnoop {}{T.~Sj{\"o}strand, S.~Mrenna, and P.~Z. Skands, ``{PYTHIA} 6.4
  physics and manual'',} \textit{ JHEP} \textbf{ 05} (2006) 026,
  \href{http://dx.doi.org/10.1088/1126-6708/2006/05/026}{\doi{10.1088/1126-6708/2006/05/026}},
  \href{http://www.arXiv.org/abs/hep-ph/0603175}{\texttt{arXiv:hep-ph/0603175}}.

\bibitem{pythia8}
T.~Sj{\"o}strand\hrefCMSnoop {}{ {et~al.}, ``{An introduction to PYTHIA
  8.2}'',} \textit{ Comput. Phys. Commun.} \textbf{ 191} (2015) 159,
  \href{http://dx.doi.org/10.1016/j.cpc.2015.01.024}{\doi{10.1016/j.cpc.2015.01.024}},
\href{http://www.arXiv.org/abs/1410.3012}{\texttt{arXiv:1410.3012}}.

\bibitem{nnpdf}
R.~D. Ball\hrefCMSnoop {}{ {et~al.}, ``{Parton distributions for the LHC run
  II}'',} \textit{ JHEP} \textbf{ 04} (2015) 040,
  \href{http://dx.doi.org/10.1007/JHEP04(2015)040}{\doi{10.1007/JHEP04(2015)040}},
  \href{http://www.arXiv.org/abs/1410.8849}{\texttt{arXiv:1410.8849}}.

\bibitem{sm_ww}
T.~Gehrmann\hrefCMSnoop {}{ {et~al.}, ``{$W^{+}W^{-}$ production at hadron
  colliders in NNLO QCD}'',} \textit{ Phys. Rev. Lett.} \textbf{ 113} (2014)
  212001,
  \href{http://dx.doi.org/10.1103/PhysRevLett.113.212001}{\doi{10.1103/PhysRevLett.113.212001}},
  \href{http://www.arXiv.org/abs/1408.5243}{\texttt{arXiv:1408.5243}}.

\bibitem{sm_ttbar}
\hrefCMSnoop {}{M.~Czakon and A.~Mitov, ``{NNLO corrections to top pair
  production at hadron colliders: the quark-gluon reaction}'',} \textit{ JHEP}
  \textbf{ 01} (2013) 080,
  \href{http://dx.doi.org/10.1007/JHEP01(2013)080}{\doi{10.1007/JHEP01(2013)080}},
  \href{http://www.arXiv.org/abs/1210.6832}{\texttt{arXiv:1210.6832}}.

\bibitem{sm_top}
\hrefCMSnoop {}{N.~Kidonakis, ``Next-to-next-to-leading-order collinear and
  soft gluon corrections for $t$-channel single top quark production'',}
  \textit{ Phys. Rev. D} \textbf{ 83} (2011) 091503,
  \href{http://dx.doi.org/10.1103/PhysRevD.83.091503}{\doi{10.1103/PhysRevD.83.091503}},
  \href{http://www.arXiv.org/abs/1103.2792}{\texttt{arXiv:1103.2792}}.

\bibitem{single_cross_sec1}
\hrefCMSnoop {}{J.~Campbell, R.~K. Ellis, and F.~Tramontano, ``{Single top
  production and decay at next-to-leading order}'',} \textit{ Phys. Rev.}
  \textbf{ 70} (2004) 094012,
  \href{http://dx.doi.org/10.1103/PhysRevD.70.094012}{\doi{10.1103/PhysRevD.70.094012}},
  \href{http://www.arXiv.org/abs/hep-ph/0408158}{\texttt{arXiv:hep-ph/0408158}}.

\bibitem{single_cross_sec2}
\hrefCMSnoop {}{G.~Panico, O.~Matsedonskyi, and A.~Wulzer, ``{On the
  interpretation of top partners searches}'',} \textit{ JHEP} \textbf{ 12}
  (2014) 097,
  \href{http://dx.doi.org/10.1007/JHEP12(2014)097}{\doi{10.1007/JHEP12(2014)097}},
  \href{http://www.arXiv.org/abs/1409.0100}{\texttt{arXiv:1409.0100}}.

\bibitem{geant4}
\hrefCMSnoop {}{{GEANT4} Collaboration, ``{GEANT4---a simulation toolkit}'',}
  \textit{ Nucl. Instrum. Meth. A} \textbf{ 506} (2003) 250,
  \href{http://dx.doi.org/10.1016/S0168-9002(03)01368-8}{\doi{10.1016/S0168-9002(03)01368-8}}.

\bibitem{particleflow}
\href {http://cdsweb.cern.ch/record/1194487}{{CMS Collaboration},
  ``Particle--flow event reconstruction in {CMS} and performance for jets,
  taus, and {\MET}'',} CMS Physics Analysis Summary CMS-PAS-PFT-09-001, 2009.

\bibitem{particleflow2}
\href {http://cdsweb.cern.ch/record/1247373}{{CMS Collaboration},
  ``Commissioning of the particle--flow event reconstruction with the first
  {LHC} collisions recorded in the {CMS} detector'',} CMS Physics Analysis
  Summary CMS-PAS-PFT-10-001, 2010.

\bibitem{CMS:electrons}
\hrefCMSnoop {}{{CMS Collaboration}, ``{Performance of electron reconstruction
  and selection with the CMS detector in proton-proton collisions at $\sqrt{s}
  = 8\TeV$}'',} \textit{ JINST} \textbf{ 10} (2015) P06005,
  \href{http://dx.doi.org/10.1088/1748-0221/10/06/P06005}{\doi{10.1088/1748-0221/10/06/P06005}},
  \href{http://www.arXiv.org/abs/1502.02701}{\texttt{arXiv:1502.02701}}.

\bibitem{CMS:muons}
\hrefCMSnoop {}{{CMS Collaboration}, ``{Description and performance of track
  and primary-vertex reconstruction with the CMS tracker}'',} \textit{ JINST}
  \textbf{ 9} (2014) P10009,
  \href{http://dx.doi.org/10.1088/1748-0221/9/10/P10009}{\doi{10.1088/1748-0221/9/10/P10009}},
  \href{http://www.arXiv.org/abs/1405.6569}{\texttt{arXiv:1405.6569}}.

\bibitem{TagAndProbe}
\hrefCMSnoop {}{{CMS Collaboration}, ``{Measurements of inclusive W and Z cross
  sections in pp collisions at $\sqrt{s} = 7\TeV$}'',} \textit{ JHEP} \textbf{
  01} (2011) 080,
  \href{http://dx.doi.org/10.1007/JHEP01(2011)080}{\doi{10.1007/JHEP01(2011)080}},
  \href{http://www.arXiv.org/abs/1012.2466}{\texttt{arXiv:1012.2466}}.

\bibitem{Cacciari:dispelling}
\hrefCMSnoop {}{M.~Cacciari and G.~P. Salam, ``{Dispelling the $N^{3}$ myth for
  the $k_t$ jet-finder}'',} \textit{ Phys. Lett. B} \textbf{ 641} (2006) 57,
  \href{http://dx.doi.org/10.1016/j.physletb.2006.08.037}{\doi{10.1016/j.physletb.2006.08.037}},
  \href{http://www.arXiv.org/abs/hep-ph/0512210}{\texttt{arXiv:hep-ph/0512210}}.

\bibitem{antikt}
\hrefCMSnoop {}{M.~Cacciari, G.~P. Salam, and G.~Soyez, ``{The anti-$k_t$ jet
  clustering algorithm}'',} \textit{ JHEP} \textbf{ 04} (2008) 063,
  \href{http://dx.doi.org/10.1088/1126-6708/2008/04/063}{\doi{10.1088/1126-6708/2008/04/063}},
  \href{http://www.arXiv.org/abs/0802.1189}{\texttt{arXiv:0802.1189}}.

\bibitem{fastjet1}
\hrefCMSnoop {}{M.~Cacciari, G.~P. Salam, and G.~Soyez, ``{The catchment area
  of jets}'',} \textit{ JHEP} \textbf{ 04} (2008) 005,
  \href{http://dx.doi.org/10.1088/1126-6708/2008/04/005}{\doi{10.1088/1126-6708/2008/04/005}},
  \href{http://www.arXiv.org/abs/0802.1188}{\texttt{arXiv:0802.1188}}.

\bibitem{fastjet2}
\hrefCMSnoop {}{M.~Cacciari and G.~P. Salam, ``{Pileup subtraction using jet
  areas}'',} \textit{ Phys. Lett. B} \textbf{ 659} (2008) 119,
  \href{http://dx.doi.org/10.1016/j.physletb.2007.09.077}{\doi{10.1016/j.physletb.2007.09.077}},
  \href{http://www.arXiv.org/abs/0707.1378}{\texttt{arXiv:0707.1378}}.

\bibitem{Chatrchyan:2011ds}
\hrefCMSnoop {}{{CMS Collaboration}, ``{Determination of jet energy calibration
  and transverse momentum resolution in CMS}'',} \textit{ JINST} \textbf{ 6}
  (2011) P11002,
  \href{http://dx.doi.org/10.1088/1748-0221/6/11/P11002}{\doi{10.1088/1748-0221/6/11/P11002}},
  \href{http://www.arXiv.org/abs/1107.4277}{\texttt{arXiv:1107.4277}}.

\bibitem{CMS-PAS-BTV-15-001}
\href {https://cds.cern.ch/record/2138504?ln=en}{{CMS Collaboration},
  ``Identification of b quark jets at the CMS experiment in the LHC Run 2'',}
  CMS Physics Analysis Summary CMS-PAS-BTV-15-001, 2015.

\bibitem{BTV_12_paper}
\hrefCMSnoop {}{{CMS Collaboration}, ``{Identification of b-quark jets with the
  CMS experiment}'',} \textit{ JINST} \textbf{ 8} (2011), no.~04, P04013,
  \href{http://dx.doi.org/10.1088/1748-0221/8/04/P04013}{\doi{10.1088/1748-0221/8/04/P04013}},
  \href{http://www.arXiv.org/abs/1211.4462}{\texttt{arXiv:1211.4462}}.

\bibitem{diff-ttbar}
\hrefCMSnoop {}{{CMS Collaboration}, ``{Measurement of the differential cross
  section for top quark pair production in pp collisions at $\sqrt{s} =
  8\TeV$}'',} \textit{ Eur. Phys. J. C} \textbf{ 75} (2015) 542,
  \href{http://dx.doi.org/10.1140/epjc/s10052-015-3709-x}{\doi{10.1140/epjc/s10052-015-3709-x}},
  \href{http://www.arXiv.org/abs/1505.04480}{\texttt{arXiv:1505.04480}}.

\bibitem{CMS_lumi}
\href {https://cds.cern.ch/record/2138682}{{CMS Collaboration}, ``CMS
  luminosity measurement for the 2015 data taking period'',} CMS Physics
  Analysis Summary CMS-PAS-LUM-15-001, 2016.

\bibitem{cms_ttbar}
\hrefCMSnoop {}{{CMS Collaboration}, ``{Measurement of the $\ttbar$ production
  cross section in the dilepton channel in pp collisions at $\sqrt{s} =
  8\TeV$}'',} \textit{ JHEP} \textbf{ 02} (2014) 024,
  \href{http://dx.doi.org/10.1007/JHEP02(2014)024}{\doi{10.1007/JHEP02(2014)024}},
  \href{http://www.arXiv.org/abs/1312.7582}{\texttt{arXiv:1312.7582}}.

\bibitem{cms_top_paper}
\hrefCMSnoop {}{{CMS Collaboration}, ``{Measurement of the inclusive cross
  section of single top-quark production in the $t$-channel at 13\TeV}'',}
  \textit{ Phys. Lett. B} , in proofs,
  \href{http://www.arXiv.org/abs/1610.00678}{\texttt{arXiv:1610.00678}}.

\bibitem{CMS-PAS-SMP-15-004}
\href {https://cds.cern.ch/record/2093537}{{CMS Collaboration}, ``Measurement
  of inclusive W and Z boson production cross sections in pp collisions at
  $\sqrt{s} = 13\TeV$'',} CMS Physics Analysis Summary CMS-PAS-SMP-15-004,
  2015.

\bibitem{WplusC}
\hrefCMSnoop {}{{CMS Collaboration}, ``{Measurement of associated W+charm
  production in pp collisions at $\sqrt{s} = 7\TeV$}'',} \textit{ JHEP}
  \textbf{ 02} (2014) 013,
  \href{http://dx.doi.org/10.1007/JHEP02(2014)013}{\doi{10.1007/JHEP02(2014)013}},
  \href{http://www.arXiv.org/abs/1310.1138}{\texttt{arXiv:1310.1138}}.

\bibitem{WplusB_paper}
\hrefCMSnoop {}{{CMS Collaboration}, ``{Measurement of the production cross
  section of the W boson in association with two b jets in pp collisions at
  $\sqrt{s} = 8\TeV$}'',} \textit{ Eur. Phys. J. C} \textbf{ 77} (2017) 92,
  \href{http://dx.doi.org/10.1140/epjc/s10052-016-4573-z}{\doi{10.1140/epjc/s10052-016-4573-z}},
  \href{http://www.arXiv.org/abs/1608.07561}{\texttt{arXiv:1608.07561}}.

\bibitem{lumi}
\hrefCMSnoop {}{{ATLAS Collaboration}, ``{Measurement of the inelastic
  proton-proton cross section at $\sqrt{s} = 13\TeV$ with the ATLAS detector at
  the LHC}'',} \textit{ Phys. Rev. Lett.} \textbf{ 117} (2016), no.~18, 182002,
  \href{http://dx.doi.org/10.1103/PhysRevLett.117.182002}{\doi{10.1103/PhysRevLett.117.182002}},
\href{http://www.arXiv.org/abs/1606.02625}{\texttt{arXiv:1606.02625}}.

\bibitem{pdf_lhc}
\hrefCMSnoop {}{J.~Butterworth {et~al.}, ``{PDF4LHC recommendations for LHC Run
  II}'',} \textit{ J. Phys. G} \textbf{ 43} (2016) 023001,
  \href{http://dx.doi.org/10.1088/0954-3899/43/2/023001}{\doi{10.1088/0954-3899/43/2/023001}},
\href{http://www.arXiv.org/abs/1510.03865}{\texttt{arXiv:1510.03865}}.

\bibitem{log_prior_ref1}
W.~T. Eadie {et~al.}, ``Statistical methods in experimental physics''.
\newblock (North Holland, Amsterdam), 1971.
\newblock [See pages 6, 79--80].

\bibitem{log_prior_ref2}
F.~James, ``Statistical methods in experimental physics''.
\newblock (World Scientific, Singapore), 2 edition, 2006.
\newblock [See pages 83--87].

\bibitem{Theta_web}
\href {http://www-ekp.physik.uni-karlsruhe.de/~ott/theta/theta-auto/}{J.~Ott,
  ``Theta -- theta-auto testing documentation'',} (2012).

\end{thebibliography}\endgroup
\cleardoublepage \appendix\section{The CMS Collaboration \label{app:collab}}\begin{sloppypar}\hyphenpenalty=5000\widowpenalty=500\clubpenalty=5000\textbf{Yerevan Physics Institute,  Yerevan,  Armenia}\\*[0pt]
A.M.~Sirunyan, A.~Tumasyan
\vskip\cmsinstskip
\textbf{Institut f\"{u}r Hochenergiephysik,  Wien,  Austria}\\*[0pt]
W.~Adam, E.~Asilar, T.~Bergauer, J.~Brandstetter, E.~Brondolin, M.~Dragicevic, J.~Er\"{o}, M.~Flechl, M.~Friedl, R.~Fr\"{u}hwirth\cmsAuthorMark{1}, V.M.~Ghete, C.~Hartl, N.~H\"{o}rmann, J.~Hrubec, M.~Jeitler\cmsAuthorMark{1}, A.~K\"{o}nig, I.~Kr\"{a}tschmer, D.~Liko, T.~Matsushita, I.~Mikulec, D.~Rabady, N.~Rad, B.~Rahbaran, H.~Rohringer, J.~Schieck\cmsAuthorMark{1}, J.~Strauss, W.~Waltenberger, C.-E.~Wulz\cmsAuthorMark{1}
% ...
\vskip\cmsinstskip
% ...
1:~~Also at Vienna University of Technology, Vienna, Austria\\
% ...

\end{sloppypar}
\end{document}